\documentclass[twocolumn,english,showpacs]{revtex4-2}
\usepackage{geometry}
\usepackage{ragged2e}
\usepackage{xcolor}
\usepackage{babel}
\usepackage{units}
\usepackage{amsmath}
\usepackage{amssymb}
\usepackage{stmaryrd}
\usepackage{graphicx}
\usepackage{wasysym}
\usepackage{bbold}
\usepackage{mathtools}
\usepackage{blkarray, bigstrut}
\usepackage{float}
\usepackage[bookmarks=false]{hyperref}
\newcommand{\ket}[1]{\vert{#1}\rangle}
\newcommand{\up}{\mid\uparrow_q\rangle}
\newcommand{\bra}[1]{\langle{#1}|}
\newcommand{\down}{\mid\downarrow_q\rangle}

\newcommand{\CM}[1]{{\color{black}#1}}

\newcommand{\DQC}{Duke Quantum Center, Departments of Electrical and Computer Engineering and Physics, \\Duke University, Durham, NC 27708}

\begin{document}
\title{Entanglement fidelity limits of photonically-networked atomic qubits \\ from recoil and timing}
\date{\today}

\author{Yichao Yu}
\email{Corresponding author: yyc1992@gmail.com}
\author{Sagnik Saha}
\author{Mikhail Shalaev}
\author{George Toh}
\author{Jameson O'Reilly}
\email{Present Address: Department of Physics, University of Oregon, Eugene, OR 97331}
\author{Isabella Goetting}
\author{Ashish Kalakuntla}
\author{Christopher Monroe}
\affiliation{\DQC}

\begin{abstract}
The remote entanglement of two atomic quantum memories through photonic interactions is accompanied by atomic momentum recoil. 
When the interactions occur at different times, such as from the random emission over the lifetime of the atomic excited state, the difference in recoil timing can expose ``which-path'' information and ultimately lead to decoherence. Time-bin encoded photonic qubits can be particularly sensitive to asynchronous recoil timing. In this paper we study the limits of entanglement fidelity in atomic systems due to recoil and other timing imbalances and show how these effects can be suppressed or even eliminated through proper experimental design.
\end{abstract}

\maketitle

High-quality and fast photonic entanglement links between good quantum memories are crucial for large-scale modular quantum computers and quantum networks \cite{Awschalom2021}. The central resource in these links is the collection and interference of photons from quantum memories \cite{Kimble2008}, typically through high numerical-aperture objectives or optical cavities. Such systems have been proposed and demonstrated in optically-active quantum memory systems such as trapped ions \cite{barrett_efficient_2005, moehring_quantum_2007, Moehring2007E, Duan2010, Hucul2015, Stephenson2020, Oreilly2024, Saha2025}, neutral atoms \cite{Ritter2012, Weinfurter2022, Li2024,Li2025}, and solid-state quantum emitters \cite{Bernien2013, lukin_siv_entanglement_2024, ruskuc2024scalable}. 

The quantum entanglement of two quantum memories through a photonic channel can be classified by the way photons carry quantum information \cite{Duan2010}. In Type I protocols \cite{Cabrillo1999, duan_long-distance_2001}, memories are entangled through the heralded detection of a single photon following interference between which of the two memories produced the photon. 
% CONSIDER REVISING
In Type II protocols, each memory becomes entangled with an intrinsic photonic qubit, and the memories become entangled either through ``pitch-and-catch" protocols \cite{Cirac1997} or heralded detection of two photons following their interference \cite{Duan2003, barrett_efficient_2005}.
%Type-II protocols are favored because their fidelity is much less sensitive to ``which path'' errors. 
Type I protocols require stringent optical path stability and also suffer from differential atomic recoil \cite{Eichman1993}. Type II protocols are much less sensitive to these effects but are generally slower, as they require two photonic interactions, each typically occurring with low probability. However, with continual gains in photonic input/output efficiency \cite{UIBK2023,Carter2024,Li2024,Li2025}, this drawback is fading.

%Pitch-and-catch protocols with strongly-coupled cavities can be treated similarly, , as the recoil can depend critically upon the cavity parameters \cite{Kikura2025}.
In this paper, we consider Type II photonic entanglement between qubit memories (represented by states $\ket{\downarrow}$ and $\ket{\uparrow}$), where photons are emitted, mode-matched onto a 50/50 beamsplitter (BS), and detected, ideally heralding entangled Bell states of the qubit memories in the form
\begin{align}
\Psi^{\pm}_{AB} = \frac{\ket{\downarrow_A \uparrow_B} \pm \ket{\uparrow_A \downarrow_B}}{\sqrt{2}}. \label{eqn:Bell}
\end{align}
We calculate the effects of recoil differences between the two memories due to different times of excitation or emission, consider the resulting entanglement fidelity limits, and discuss how to suppress or even eliminate these errors. 
This analysis can also apply to pitch-and-catch schemes \cite{Cirac1997, Ritter2012} and atom-cavity systems with high photon output rates \cite{Li2024}. This paper relates to the original discovery and observation of these effects \cite{Saha2025} and subsequent theoretical work \cite{Kikura2025, Apolin2025, VanDamnitAll2025}.

\begin{figure*}[t!]%
\centering
\includegraphics[width=0.95\textwidth]{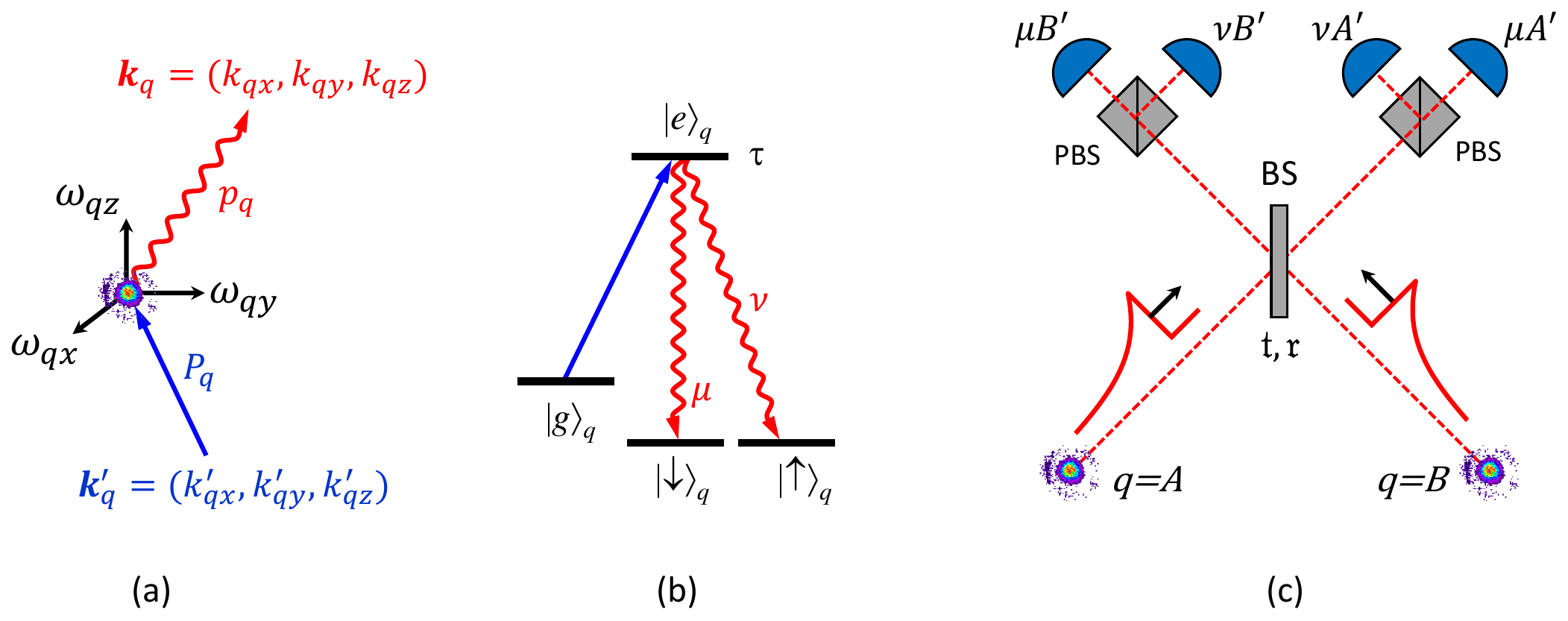}
\caption{Schematics of photon generation, interference and detection. 
(a) Atom-like emitter $q$ is bound in 3D with harmonic frequencies $\omega_{qi}$. (Extension to normal modes of a chain of atoms is straighforward and discussed in the text.) Atom is excited with probability $P_q$ by a laser pulse having wavevector $\mathbf{k}'_q$ and emits a single photon with wavevector $\mathbf{k}_q$ with probability $p_q$.
(b) A laser pulse (blue) excites atom $q$ from state $\ket{g}_q$ to $\ket{e}_q$ (radiative lifetime $\tau$), followed by emission (red) to an equal superposition of atomic qubit states $\down$ and $\up$, correlated with photonic qubit state $\mu$ and $\nu$, respectively.
(c) Photons from atomic qubits $q=A$ and $q=B$ are mode-matched onto the input ports of a beamsplitter (BS) that has field transmission and reflection coefficients $\mathfrak{t}$ and $\mathfrak{r}$ with $\mathfrak{t}^2+\mathfrak{r}^2=1$. After interference, the BS output ports are relabeled by $q' \in A',B'$ as shown.
Finally, the photon qubit states are separated and detected. The figure shows an example for polarization qubits, where additional polarizing beamsplitters (PBS) direct each qubit state to independent detectors labeled by the photonic qubit $\mu$ or $\nu$ and output spatial mode $A'$ or $B'$.}
%(d) Time-domain photonic wavepackets $f_q(t)$ from emitters $q \in A,B$ at the beamsplitter, including the possibility of an imbalance from excitation times $t_A^0$ and $t_B^0$ or path lengths $l_A$ and $l_B$ between the emitters and the BS.}
\label{fig:photons}
\end{figure*}
In practice, the entanglement fidelity relies upon the indistinguishability of the photons, requiring transverse mode-matching of photons on a coupling BS, longitudinal matching of photons through their timing and bandwidth, and synchronization of quantum memory excitation. Any misalignment or imbalance can reduce the ultimate entanglement fidelity based on the spatial and temporal overlap of the photonic wavefunctions. 
Photonic time jitter can result from imperfect photonic elements such as excitation time jitter ($<100$ ps for ultrafast pulsed excitation), photon detector jitter (10-100 ps \cite{APDjitter1996}), or vibratory noise on optical elements ($<10$ fs for $\sim 1$ $\mu$m displacements). All of the above sources of mismatch or timing noise are either negligible or technical in nature and can be reduced to insignificance.

However, the emission/detection process is intrinsically random in time, typically distributed over 1-100 ns. This can degrade entanglement coherence through the entanglement with atomic motion from photon recoil \cite{Saha2025, Kikura2025,Apolin2025, VanDamnitAll2025}.  
%The finite radiative lifetime of atomic emitters results in an inherent spread in time of the photonic wavepacket, typically in the realm of nanoseconds. 
Ultimately, the detection of the two photons after the BS occurs at random times, resulting in a fundamental source of entanglement decoherence that may dominate other technical sources of noise.

%********************************************************************
\subsection*{\CM{Photonic Wavepacket Description}}
In Fig. \ref{fig:photons}, we show the schematic of the Type-II quantum memory entanglement scheme. Two atomic emitters $q \in (A,B)$ with position operator $\hat{\mathbf{r}}_q$ from equilibrium are deterministically excited with fast optical pulses of wavevector $\mathbf{k}'_q$ at time $t=0$. (Any relative delays in excitation times will be treated later.) With probability $P_q\sim1$, each atom is excited from its ground state $\ket{g}$ to the short-lived excited state $\ket{e}_q$ having radiative lifetime $\tau$ typically of order $10^{-8}$ s.
The emitted photons are collected with probability $p_q$ from each atom and with wavevector $\mathbf{k}_q$ are directed to a BS. The collection efficiency includes the solid angle of collection optics and all associated losses leading up to the BS. \CM{We note that Ref. \cite{Apolin2025} generalizes to the case of emission into large solid angles where the recoil can also acquire transverse components.}

The photons \CM{each encode a} qubit labeled by the index $\pi \in (\mu,\nu)$. These two values could be two orthogonal polarization states, resolvable frequency states, photonic orbital angular momentum states, or time-bin encoded states, for example. Any fractional difference in the wavenumbers of the photon qubit states $\mu$ and $\nu$ are assumed negligible.
After emission and successful photon collection, the atom-photon system is in an equal superposition of $\down\ket{\mu}$ and $\up\ket{\nu}$, as shown in Fig. \ref{fig:photons}b. (Unequal branching ratios can also be considered, affecting only the ultimate rate of Bell state creation.)

The bandwidth $1/\tau$ of the photons is much smaller than their center (optical) frequency $\omega_0$, so we can safely write the photon field as propagating at a single frequency and wavenumber $|\mathbf{k}_q|=\omega_0/c$.
We write the field operators for the creation/annihilation of a photon in qubit mode $\pi$ and spatial mode $q$ as
\begin{align} \label{eqn:adag}
a_{\pi q}^{\dag} &= \int dt' f_q(t') a_{\pi q}^{\dag}(t') \\
a_{\pi q} &= \int dt' f_q(t') a_{\pi q}(t'),
\end{align}
where
\begin{align}
f_q(t)
=\frac{1}{\sqrt{\tau}}e^{-t/2\tau}%\Theta(t-t_q^0)
%\exp\left(-\frac{t-t^0_q}{2\tau}\right)
\label{eqn:wave}
\end{align}
is the exponentially decaying wavepacket field envelope defined for $t\ge 0$, independent of the photonic qubit state $\pi$.
%, and $\Theta(t)$ is the Heaviside function. 
The probability of the photon being emitted at time $t$ is $f_q(t)^2 dt$, with ${\int f_q(t)^2dt =1}$. 
The operators $a_{\pi q}^{\dag}(t)$ and $a_{\pi q}(t)$ create/annihilate a single photon in qubit mode $\pi$ and spatial mode $q$ at time $t$ and satisfy the commutation relation ${[a_{\pi q}(t_1),a_{\pi q}^{\dag}(t_2)]=\delta(t_1-t_2)}$. 

%****************************************************
\subsection*{Photon Generation}
We now evolve the quantum state of the entire system composed of atomic states spanning $\ket{g}_q$, $\ket{e}_q$, $\up$, and $\down$; the two modes of the each emitted photon field spanning $\ket{n_{\mu q} n_{\nu q}}$ with $n_{\mu q}$ and $n_{\nu q}$ representing the photon occupation numbers in modes ${\mu}$ and ${\nu}$ for each atom; and the motion of each atom. 

The initial state of each atomic system is
\begin{align}
%\rho_q &=\ket{g}_q \bra{g}_q \otimes \ket{0_H0_V}_q \bra{0_H0_V}_q\otimes M_{q},
\ket{\psi^{(0)}_q} = \ket{g}_q \otimes \ket{0_{\mu q}0_{\nu q}} \otimes\ket{\boldsymbol\alpha_{q}}, \label{eqn:initial}
\end{align}
where $\ket{\boldsymbol\alpha_{q}}=\bigotimes_i \ket{\alpha_{qi}}$ is the initial state of motional mode $i$, each taken to be a pure coherent state \cite{Glauber1963} of amplitude $\alpha_{qi}$. (Later, we will generalize to thermal states of motion \CM{by forming an appropriate mixture of coherent states}.) For atom emitters that are one of $N$ coupled atoms such as a crystal of trapped ions, the mode $i$ is one of the $3N$ normal modes of motion.

After excitation but before photon interference on the BS, the state at time $t$ evolves to an entangled state between photonic, atomic and motional degrees of freedom. Given that a photon is produced and collected, the resulting state of each atom and photon is written,
%\begin{align} \label{eqn:stateb4BS}
%\ket{\psi^{(1)}_q(t)} =&\sqrt{P_qp_q}\int_0^\infty dt' f(t') e^{-iH_q (t-t')}\\
%\times & \frac{1}{\sqrt{2}}\left[\hat{\Gamma}^q_{\downarrow\mu}(t')+\hat{\Gamma}^q_{\uparrow\nu}(t')\right] e^{-iH_q t'} \hat{E}^q \ket{\psi_q^{(0)}}. \nonumber 
%\end{align}
\begin{align} \label{eqn:stateb4BS}
\ket{\psi^{(1)}_q(t)} &=\sqrt{P_qp_q}\int_0^t dt' f(t')  \\
\times \frac{1}{\sqrt{2}}&\left[\hat{\Gamma}^q_{\downarrow\mu}(t')  \hat{E}^q(0) 
+ \hat{\Gamma}^q_{\uparrow\nu}(T+t') \hat{E}^q(T)\right]
%&\times \frac{1}{\sqrt{2}}\left[\hat{\Gamma}^q_{\downarrow\mu}(t') e^{-iH_q t'} \hat{E}^q(t^0_q) + \right. \\
%      &\left. \hspace{2.3mm}e^{iH_q T}\hat{\Gamma}^q_{\uparrow\nu}(t') e^{-iH_q t'} \hat{E}^q(t^0_q) e^{-iH_q T} \right]
\ket{\psi_q^{(0)}}, \nonumber
\end{align}
where we adopt the convention of normalizing conditional states to the probability of their occurance.
The atomic excitation and emission operators are
\begin{align} \label{eqn:excitation}
\hat{E}^q(t) &= e^{i\mathbf{k}'_q \cdot \mathbf{\hat{r}}_q(t)}
\ket{e_q}\bra{g_q} \\
\hat{\Gamma}^q_{S\pi}(t) &= e^{-i\mathbf{k}_q \cdot \mathbf{\hat{r}}_q(t)}
\ket{S_q}\bra{e_q} a^\dagger_{\pi q}(t), \label{eqn:emission}
\end{align}
where we have included the explicit time dependence in the atomic position operator $\mathbf{\hat{r}}_q(t)$ by transforming to rotating frames of each mode $i$ under the Hamiltonian ${H_q=\sum_{i}\omega_{qi}(n_{qi}+\tfrac12)}$ with harmonic frequency $\omega_{qi}$ and phonon occupation number $n_{qi}$ in each dimension/mode~$i$.
%For collective normal modes, the position is expressed as a sum over the $N$ normal mode coordinate operators in each of the 3 dimensions, $\hat{r}_{qi}(t)=\sum_{i_m=1}^{N} b_{q i_m} \hat{r}_{q i_m}(t)$, where the normal mode transformation matrix satisfies $\sum_q b_{q i_m}b_{q i_{m'}}=\delta_{i_m i_{m'}}$ and $\sum_{i_m} b_{q i_m}b_{q' i_m}=\delta_{qq'}$ \cite{James1998}.

The evolution in Eq. \ref{eqn:stateb4BS} results in a superposition of (a) a single photon produced at time $t'$ with photonic qubit state $\ket{\mu}$ correlated with qubit state $\down$, and (b) a single photon produced at time $T+t'$ with photonic qubit state $\ket{\nu}$ correlated with qubit state ${\up}$. We account for time-bin encoding by setting a fixed delay $T\gg \tau$ such that the two time bins do not overlap, during which time the atomic memory qubits states are swapped \cite{Bernien2013}. For other types of photonic qubit encodings such as polarization, we simply set $T=0$.
The state in Eq. \ref{eqn:stateb4BS} is a superposition over all potential emission times $t'$ weighted by the photon wavepacket amplitude $f(t')$.

The momentum recoil operators appearing in Eqs. \ref{eqn:excitation}-\ref{eqn:emission} are
\begin{align}
e^{\pm ik_{qi} \hat{r}_{qi}(t)}\ket{\alpha_{qi}}
&=\mathcal{D}\!\left[\pm i\eta_{qi}e^{i\omega_{qi}t}\right]\ket{\alpha_{qi}}
%&=e^{\pm i\eta_{qi} \text{Re}(\alpha_{qi}e^{-i\omega_{qi}t})}\ket{\alpha_{qi} \pm i\eta_{qi}e^{i\omega_{qi}t}} \nonumber 
%e^{-ik_{qi} r_{qi}(t)}\ket{\alpha_{qi}}
%&=\mathcal{D}(-i\eta_{qi}e^{i\omega_{qi}t})\ket{\alpha_{qi}} \\
%&=e^{-i\eta_{qi} \text{Re}(\alpha_{qi}e^{-i\omega_{qi}t})}\ket{\alpha_{qi}-i\eta_{qi}e^{i\omega_{qi}t}}. \nonumber
%&\text{Evolution:} &e^{-iH_{qi}t}\ket{\alpha_{qi}}
%&=\ket{\alpha_{qi} e^{-i\omega_{qi} t}}.
\end{align}
Here, the coherent displacement operator \cite{Glauber1963} ${\mathcal{D}[\xi]\ket{\alpha}=e^{i\text{Im}\xi\alpha^*}\ket{\alpha+\xi}}$ 
and the Lamb-Dicke parameters for excitation ${\eta_{qi}' = k_{qi}'b_{q i}\sqrt{\hbar/2m\omega_{qi}}}$ and emission ${\eta_{qi} = k_{qi}b_{q i}\sqrt{\hbar/2m\omega_{qi}}}$ characterize the $i^{th}$ component of recoil of atom $q$ having mass $m$. The normal mode matrix $b_{qi}$ describes the participation of atom $q$ to normal mode $i$ and scales as $1/\sqrt{N}$ \cite{James1998}.

\begin{figure*}[t!]%
\centering
\includegraphics[width=0.9\textwidth]{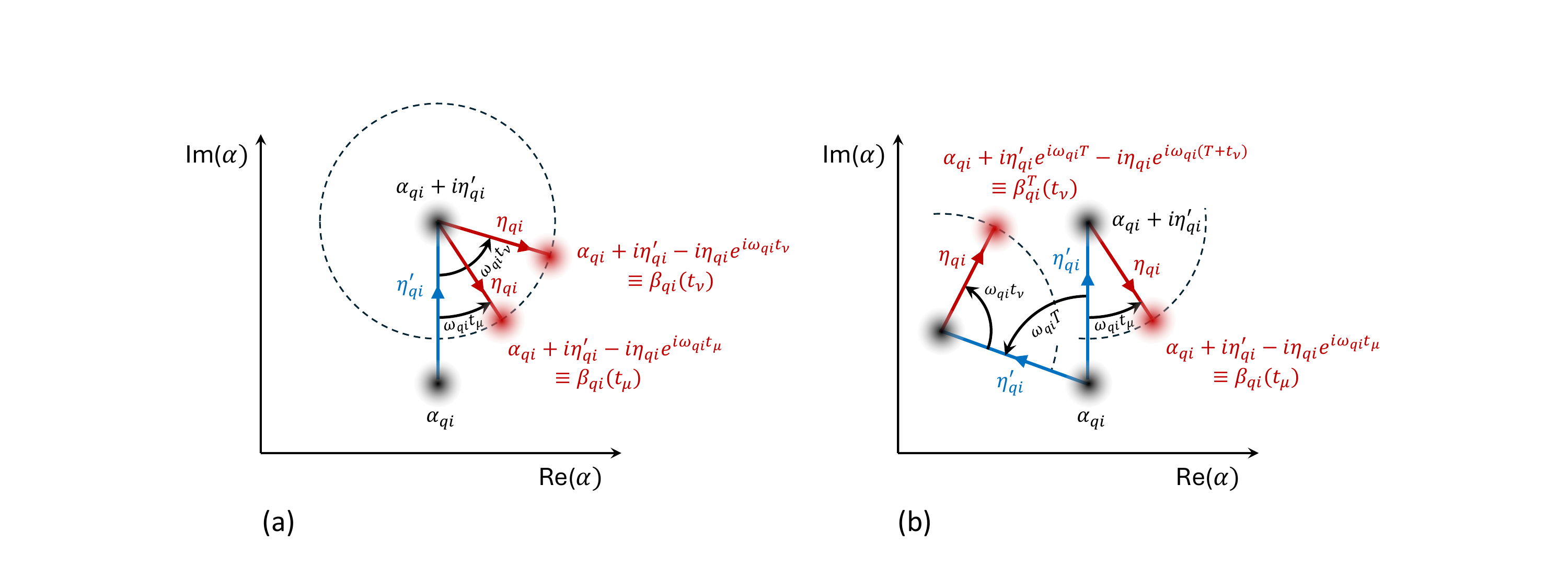}
\caption{Phase space recoil displacements of a single harmonic motional mode $i$ of atom $q$ in a frame rotating at frequency $\omega_{qi}$, with initial coherent state amplitude $\alpha_{qi}$ (lowest black dots in each plot). (a) excitation and recoil with scaled momentum kick of magnitude $\eta'_{qi}$ (blue line), then after time $t_\mu$ and $t_\nu$ emission and recoil (red lines) to coherent states $\beta_{qi}(t_\mu)$ and $\beta_{qi}(t_\nu)$ (red dots) given by Eq. \ref{eqn:alpha1}. (b) For time-bin encoding with preset delay $T$ between excitation pulses, \CM{there is additional evolution. In this case,} the excitation recoils have different phases (blue lines), corresponding to final coherent states (red dots) $\beta_{qi}(t_\mu)$ and $\beta^T_{qi}(t_\nu)$ from Eqs. \ref{eqn:alpha1}-\ref{eqn:alpha2}.}
\label{fig:phasespace}
\end{figure*}
We now rewrite Eq. \ref{eqn:stateb4BS} as
\begin{align} \label{eqn:phi1}
&\ket{\psi^{(1)}_q(t)} =\sqrt{\frac{P_q p_q}{2}}\int_{0}^t dt' f(t') \\
&\times\Bigl[\down \ket{\boldsymbol{\beta}_{q}(t')}a_{\mu q}^\dag (t') 
+ \up \ket{\boldsymbol{\beta}^T_{q}(t')}a_{\nu q}^\dag (t') \Bigr] \ket{0_{\mu q} 0_{\nu q}}, \nonumber
\end{align}
where the evolved motional coherent states of atom $q$, $\ket{\boldsymbol\beta_{q}(t')}=\bigotimes_i \ket{\beta_{qi}(t')}$ and \CM{for the case of time-bin qubit encoding there is a second coherent state ${\ket{\boldsymbol\beta^T_{q}(t')}=\bigotimes_i \ket{\beta^T_{qi}(t')}}$. The components of these coherent states are thus written}
\begin{align}
% \ket{\boldsymbol{\beta}_{q}(t')} &=
% \bigotimes_i \ket{\beta_{qi}(t')} \text{   and}\\
% \ket{\boldsymbol{\beta}^T_{q}(t')} &=
% \bigotimes_i \ket{\beta^T_{qi}(t')} \text{, where}\\
% \ket{\beta_{qi}(t')} &= e^{i\phi_{qi}(t')}\left.\Big|\alpha_{qi}+i(\eta'_{qi}-\eta_{qi}e^{i\omega_{qi}t'})\right\rangle \\
% \ket{\beta_{qi}^T(t')} &= e^{i\phi^T_{qi}(t')}\left.\Big|\alpha_{qi}+i(\eta'_{qi}-\eta_{qi}e^{i\omega_{qi}t'})e^{i\omega_{qi}T}\right\rangle ,
 \ket{\beta_{qi}(t')} &= \mathcal{D}[-i\eta_{qi}e^{i\omega_{qi}t'}]\mathcal{D}[i\eta'_{qi}]\left|\alpha_{qi}\right\rangle \label{eqn:alpha1} \\
 \ket{\beta_{qi}^T(t')} &= \mathcal{D}[-i\eta_{qi}e^{i\omega_{qi}(t'+T)}]\mathcal{D}[i\eta'_{qi}e^{i\omega_{qi}T}]\left|\alpha_{qi}\right\rangle  \label{eqn:alpha2}.
\end{align}
% with phases
% \begin{align} \label{eqn:phase}
% \phi_{qi}(t') &=   \text{Re}\left\{ \eta'_{qi}\alpha_{qi}-\eta_{qi}(\alpha_{qi}+i\eta'_{qi})e^{-i\omega_{qi}t'} \right\} \\
% \phi^T_{qi}(t') &=   \text{Re}\left\{ \eta'_{qi}\alpha_{qi}e^{-i\omega_{qi}T} \right. \label{eqn:phase2} \\ 
% &\hspace{21mm}\left.-\eta_{qi}(\alpha_{qi}e^{-i\omega_{qi}T}+i\eta'_{qi})e^{-i\omega_{qi}t'} \right\}. \nonumber
% \end{align}
\CM{A diagram of these coherent state displacements is shown in Fig. \ref{fig:phasespace}. Note that if there is no time-bin encoding, $\beta^T_{qi}(t')=\beta_{qi}(t')$.}

%****************************************************
\subsection*{Photon Interactions on Beamsplitter}

The combined photonic states from each atomic emitter are next mode-matched on the input ports of a nominal 50/50 beamsplitter (BS) (see Fig \ref{fig:photons}c).  
%Any imbalance in the path lengths $l_q$ from each emitter to the BS can be absorbed into the excitation time $\tilde{T}_{qi}_q^0 = t_q^0 + l_q/c$.
The BS transforms each time mode independently, and the photon field operators evolve according to
\begin{align}
a^\dag_{\pi A}(t) &\rightarrow \mathfrak{t} \; b^\dag_{\pi A'}(t) + \mathfrak{r} \;b^\dag_{\pi B'}(t) \\
a^\dag_{\pi B}(t) &\rightarrow \mathfrak{r} \;b^\dag_{\pi A'}(t) -\mathfrak{t} \; b^\dag_{\pi B'}(t). 
\end{align}
%\CM{ADD MODE-MISMATCHING TERMS AND LOSS TO ABOVE? (IT'S A MESS, WITH TWO MORE OUTPUT MODES)}
Here, the photon creation/annihilation operators $b^\dag_{\pi q'}(t)$ and $b_{\pi q'}(t)$ correspond to photons after the BS and satisfy ${[b_{\pi q'}(t'),b^\dag_{\pi q'}(t'')]=\delta(t'-t'')}$. The BS transmission and reflection coefficients $\mathfrak{t}$ and $\mathfrak{r}$ are taken to be real, both nominally $1/\sqrt{2}$ and satisfying $\mathfrak{r}^2+\mathfrak{t}^2=1$ (no loss). We relabel the spatial mode indices from $A,B$ to $A',B'$, as shown in Fig. \ref{fig:photons}c. 

The state of Eq. \ref{eqn:phi1} evolves according to the BS transformations above, abbreviating the functional dependences on times $t'$ and $t''$ by simply appending prime and double-prime superscripts to the photon field operators.  We find that the complete quantum state of both atoms, their motion, and the two photonic modes is 
\begin{widetext}
\begin{align} \label{eqn:completestate}
\ket{\psi^{(2)}_{AB}}
= \frac{1}{2}&\sqrt{P_A P_B p_A p_B}\int_0^\infty dt' \int_0^\infty dt'' f_A(t')f_B(t'')  \\
\times \biggl\{ 
&\ket{\downarrow_A\downarrow_B}\ket{\boldsymbol{\beta}_{A}(t')}\ket{\boldsymbol{\beta}_{B}(t'')}\hspace{.1mm}
\Big(-\mathfrak{t}^2 b'^\dag_{\mu A'} \; b''^\dag_{\mu B'} \;
+ \mathfrak{r}^2 b'^\dag_{\mu B'} \; b''^\dag_{\mu A'}
+\mathfrak{r}\mathfrak{t}\; b''^\dag_{\mu A'} \; b''^\dag_{\mu A'} -\mathfrak{r}\mathfrak{t}\; b'^\dag_{\mu B'} \; b''^\dag_{\mu B'}\Big) \biggr.
\nonumber \\
+&\ket{\uparrow_A\uparrow_B}\ket{\boldsymbol{\beta}^T_{A}(t')}\ket{\boldsymbol{\beta}^T_{B}(t'')}
\Big(-\mathfrak{t}^2 b'^\dag_{\nu A'} \; b''^\dag_{\nu B'} \;
+\mathfrak{r}^2 b'^\dag_{\nu B'} \; b''^\dag_{\nu A'}
+\mathfrak{r}\mathfrak{t}\; b'^\dag_{\nu A'} \; b''^\dag_{\nu A'} -\mathfrak{r}\mathfrak{t}\; b'^\dag_{\nu B'} \; b''^\dag_{\nu B'}\Big) \nonumber \\
+&\ket{\downarrow_A\uparrow_B}\ket{\boldsymbol{\beta}_{A}(t')}\ket{\boldsymbol{\beta}^T_{B}(t'')}
\Big(-\mathfrak{t}^2  b'^\dag_{\mu A'} \; b''^\dag_{\nu B'} \;
+ \mathfrak{r}^2 b'^\dag_{\mu B'} \; b''^\dag_{\nu A'}
+\mathfrak{r}\mathfrak{t}\; b'^\dag_{\mu A'} \; b''^\dag_{\nu A'} -\mathfrak{r}\mathfrak{t}\; b'^\dag_{\mu B'} \; b''^\dag_{\nu B'}\Big) \nonumber \\
+&\ket{\uparrow_A\downarrow_B}\ket{\boldsymbol{\beta}^T_{A}(t')}\ket{\boldsymbol{\beta}_{B}(t'')}
\Big(-\mathfrak{t}^2  b'^\dag_{\nu A'} \; b''^\dag_{\mu B'} \;
+ \mathfrak{r}^2 b'^\dag_{\nu B'} \; b''^\dag_{\mu A'}
+\mathfrak{r}\mathfrak{t}\; b'^\dag_{\nu A'} \; b''^\dag_{\mu A'} -\mathfrak{r}\mathfrak{t}\; b'^\dag_{\nu B'} \; b''^\dag_{\mu B'}\Big) \biggr\} \nonumber
\ket{0_{\mu A'} 0_{\nu A'}} \ket{0_{\mu B'} 0_{\nu B'}}.  \nonumber
\end{align}
%\CM{remove the first 8 terms?}
%****************************************************
\subsection*{Photon Measurement}
Next the two photons are measured in coincidence, heralding an entangled state of the atomic qubits. 
The first 8 of the above 16 photonic terms \CM{in Eq. \ref{eqn:completestate}}, associated with atomic qubit states $\ket{\downarrow_A\downarrow_B}$ and $\ket{\uparrow_A\uparrow_B}$, involve photons in the same qubit state $\mu$ or $\nu$ and do not project entangled states of the atomic qubits, so these events are simply discarded. The last 8 terms contain two-photon correlators that project an entangled superposition between states $\ket{\downarrow_A\uparrow_B}$ and $\ket{\uparrow_A\downarrow_B}$. There are four unique states of the two photons to consider following this coincidence measurement: either two detectors fire on opposite sides of the BS ($\ket{1001}$ or $\ket{0110}$); or they fire on the same side ($\ket{1100}$ or $\ket{0011}$). Here we have abbreviated the photonic state without subscripts; e.g., $\ket{1_{\mu A'} 0_{\nu A'}} \ket{0_{\mu B'} 1_{\nu B'}} \rightarrow \ket{1001}$. 

We model the photon measurement step by projecting one photon in state $\mu$ at time $t_\mu$ over interval $dt_\mu$, and a second photon in state $\nu$ at time $t_\nu$ over interval $dt_\nu$. The projection operator has the form
\begin{align}
%\hat{\Pi}_{\pi q'_\pi}= b^\dag_{\pi q'_\pi}(t_\pi)\ket{0_{\pi q'_\mu}} \bra{0_{\pi q'_\pi}}\; b_{\pi q'_\pi}(t_\mu),
\hat{\Pi}_{\mu q'_\mu, \nu q'_\nu}&= 
b^\dag_{\mu q'_\mu}(t_\mu)b^\dag_{\nu q'_\nu}(t_\nu)
\ket{0_{\mu q'_\mu}0_{\nu q'_\nu}} \bra{0_{\mu q'_\mu}0_{\nu 'q'_\nu}}\; 
b_{\mu q'_\mu}(t_\mu)b_{\nu q'_\nu}(t_\nu),
\end{align}
where $q'_\pi \in A',B'$ is the photonic spatial output mode associated with photonic qubit state $\pi$ for the particular correlation measured.
When the appropriate two-photon projector is applied to the state in Eq. \ref{eqn:completestate}, the integrals over $t'$ and $t''$ collapse, as $b_{\pi q}(t_1) b^\dag_{\pi q}(t_2)\ket{0_{\pi q}}=
%[b_{\pi q}(t_1),b^\dag_{\pi q}(t_2)]\ket{0_{\pi q}} =
\delta(t_1-t_2)\ket{0_{\pi q}}$. We find that one of the four pure states below is probabilistically created, labeled by 
$\ket{\psi^{(3)}_{\gamma}}=\bra{\gamma} \hat{\Pi}_{\mu q_\mu, \nu q_\nu} \ket{\psi^{(2)}_{AB}}$, where the parameter $\gamma = 1001,\ 0110,\ 1100,\ \text{or }0011$ describes which photon pair state has been measured:
\begin{align}
\ket{\psi^{(3)}_{1001}}
&= \sqrt{\chi dt_\mu dt_\nu}\left[-\mathfrak{t}^2 f_A(t_\mu)f_B(t_\nu)\ket{\downarrow_A\uparrow_B}\ket{\boldsymbol{\beta}_{A}(t_\mu)}\ket{\boldsymbol{\beta}^T_{B}(t_\nu)} 
+ \mathfrak{r}^2 f_A(t_\nu)f_B(t_\mu)\ket{\uparrow_A\downarrow_B} \ket{\boldsymbol{\beta}^T_{A}(t_\nu)}\ket{\boldsymbol{\beta}_{B}(t_\mu)}\right]\label{eqn:state1} \\
\ket{\psi^{(3)}_{0110}}
&= \sqrt{\chi dt_\mu dt_\nu}\left[\hspace{2.4mm}\mathfrak{r}^2 f_A(t_\mu)f_B(t_\nu)\ket{\downarrow_A\uparrow_B}\ket{\boldsymbol{\beta}_{A}(t_\mu)}\ket{\boldsymbol{\beta}^T_{B}(t_\nu)} 
- \mathfrak{t}^2 f_A(t_\nu)f_B(t_\mu)\ket{\uparrow_A\downarrow_B} \ket{\boldsymbol{\beta}^T_{A}(t_\nu)}\ket{\boldsymbol{\beta}_{B}(t_\mu)}\right]\label{eqn:state2} \\
\ket{\psi^{(3)}_{1100}}
&= \sqrt{\chi dt_\mu dt_\nu}\left[ \phantom{-} \mathfrak{r}\mathfrak{t} f_A(t_\mu)f_B(t_\nu)\ket{\downarrow_A\uparrow_B}\ket{\boldsymbol{\beta}_{A}(t_\mu)}\ket{\boldsymbol{\beta}^T_{B}(t_\nu)} 
+ \mathfrak{r}\mathfrak{t} f_A(t_\nu)f_B(t_\mu)\ket{\uparrow_A\downarrow_B} \ket{\boldsymbol{\beta}^T_{A}(t_\nu)}\ket{\boldsymbol{\beta}_{B}(t_\mu)}\right] \label{eqn:state3}\\
\ket{\psi^{(3)}_{0011}}
&= \sqrt{\chi dt_\mu dt_\nu}\left[ -\mathfrak{r}\mathfrak{t} f_A(t_\mu)f_B(t_\nu)\ket{\downarrow_A\uparrow_B}\ket{\boldsymbol{\beta}_{A}(t_\mu)}\ket{\boldsymbol{\beta}^T_{B}(t_\nu)} 
- \mathfrak{r}\mathfrak{t} f_A(t_\nu)f_B(t_\mu)\ket{\uparrow_A\downarrow_B} \ket{\boldsymbol{\beta}^T_{A}(t_\nu)}\ket{\boldsymbol{\beta}_{B}(t_\mu)}\right]. \label{eqn:state4}
\end{align}
\end{widetext}
Here, the normalization constant $\chi=P_AP_Bp_Ap_B\epsilon_D^2/4$, where $\epsilon_D$ is the photon detection efficiency of each detector.

%**********************************************************
\subsection*{Model of the State Fidelity and Yield}
In order to calculate the expected fidelity and success probability (yield) of the above states, we form the density matrix in each case above and trace over the motional degrees of freedom:
 \begin{align}
\rho_\gamma &= \frac{1}{\mathcal{P}_\gamma}\text{Tr}_{M} \left\{ \ket{\psi^{(3)}_{\gamma}}\bra{\psi^{(3)}_{\gamma}}\right\} \label{eqn:trace}\\
&= \Pi_{\gamma}^{\downarrow\uparrow}\hspace{0.5mm} \ket{\downarrow_A\uparrow_B}\bra{\downarrow_A\uparrow_B} + \Pi_{\gamma}^{\uparrow\downarrow}\hspace{1mm}\ket{\uparrow_A\downarrow_B}\bra{\uparrow_A\downarrow_B} \nonumber \\
&+ \frac{1}{2}C_{\gamma} \ket{\downarrow_A\uparrow_B}\bra{\uparrow_A\downarrow_B} 
+ \frac{1}{2}C_{\gamma}^*\ket{\uparrow_A\downarrow_B}\bra{\downarrow_A\uparrow_B}.
 \end{align}
We can then write the Bell state fidelity \cite{Sackett2000}
\begin{align} \label{eqn:F}
F_\gamma=\frac{1}{2}(\Pi_{\gamma}^{\downarrow\uparrow}+\Pi_{\gamma}^{\uparrow\downarrow}) + \frac{1}{2}|C_{\gamma}|=\frac{1+|C_{\gamma}|}{2},
\end{align}
where the absolute value of the coherence accounts for the phase of the expected Bell state and the populations $\Pi_{\gamma}^{\downarrow\uparrow}+\Pi_{\gamma}^{\uparrow\downarrow} = 1$.

\begin{figure}[b!]%
\centering
\includegraphics[width=0.4\textwidth]{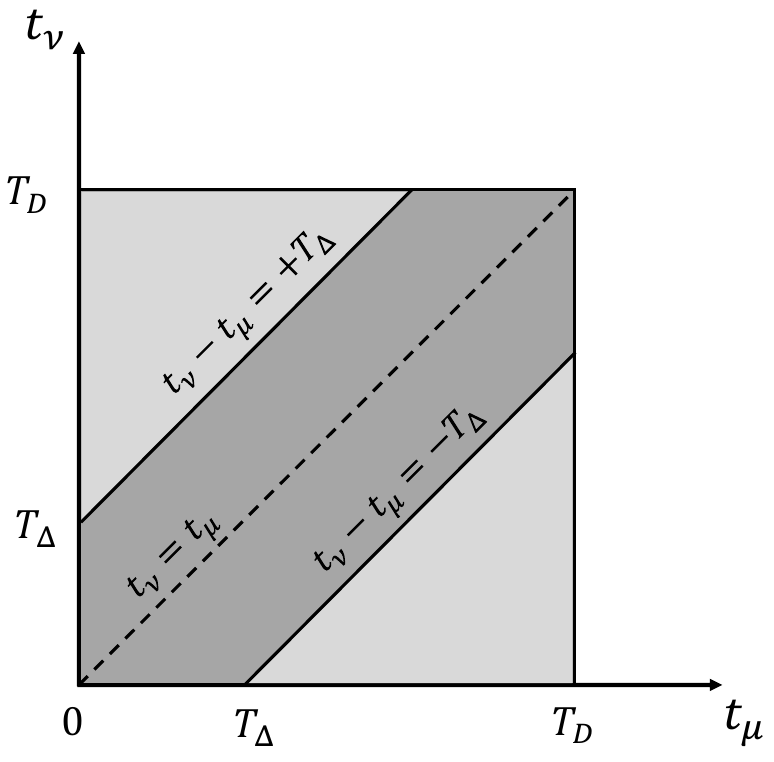}
\caption{Integration time window (dark shaded region) of two-photon events appearing on the $\mu$ and $\nu$ detectors at times $t_\mu$ and $t_\nu$, respectively.  Each detector has an independent time window of $(0,T_D)$, and also the time difference between detections is restricted to be $|t_\nu - t_\mu| < T_{\scriptscriptstyle{\Delta}}$.}
\label{fig:integ}
\end{figure}

The probability of heralding state $\ket{\psi^{(3)}_{\gamma}}$ is ${\mathcal{P}_\gamma=\text{Tr}  \ket{\psi^{(3)}_{\gamma}}\bra{\psi^{(3)}_{\gamma}}}$, 
%integrated over the detection time windows,
\begin{align}
\mathcal{P}^{-}\equiv\mathcal{P}_{1001} &= \mathcal{P}_{0110}= \chi (\mathfrak{t}^4 + \mathfrak{r}^4 )Y \label{eqn:Popp}\\
\mathcal{P}^{+}\equiv\mathcal{P}_{1100} &= \mathcal{P}_{0011}=2\chi \mathfrak{t}^2 \mathfrak{r}^2 Y \label{eqn:Psame}
\end{align}
In these expressions, the yield $Y$ of two-photon events is integrated over the joint time window $W$ of the relevant detectors over the interval $(0,T_D)$, and also over a symmetric window of time difference $|t_\nu - t_\mu|<T_{\scriptscriptstyle{\Delta}}$, as depicted by the shaded region in Fig. \ref{fig:integ}:
\begin{align}
Y &=  \iint_W dt_\mu dt_\nu f_A(t_\mu)^2 f_B(t_\nu)^2 \\
%   =  \tfrac{1}{2}\iint_W dt_{\scriptscriptstyle{\Sigma}} dt_{\scriptscriptstyle{\Delta}} \frac{e^{-t_{\scriptscriptstyle{\Sigma}}/\tau}}{\tau^2} \\
& = 1-e^{-T_{\scriptscriptstyle{\Delta}}/\tau}-e^{-(2T_D-T_{\scriptscriptstyle{\Delta}})/\tau}+e^{-2T_D/\tau}.
%= [1-e^{-(T_D-t_A^0)/\tau}][1-e^{-(T_D-t_B^0)/\tau}].
\end{align}
With no detection time-difference window ($T_{\scriptscriptstyle{\Delta}} = T_D$), the yield is $Y = (1-e^{-T_D/\tau})^2$. Henceforth, we shall take the time-difference window to be much more restrictive than the detection window, or $T_{\scriptscriptstyle{\Delta}} \ll T_D$, and the yield becomes $Y = 1-e^{-T_{\scriptscriptstyle{\Delta}}/\tau}$.

If there is a known imbalance $\delta t_0$ from emitter excitation times or optical path lengths before the BS, there is a possibility of one photon interacting with the BS before the other photon can be emitted, thereby carrying which-path information on the source of the early photon. However, such events can be simply vetoed by appropriately setting each detector window to start counting after the leading edge of the later photon wavepacket would reach the detector. For a detector time-difference window $T_{\scriptscriptstyle{\Delta}} \ll T_D$, the yield is reduced to $=e^{-\delta t_0/\tau}(1-e^{-T_{\scriptscriptstyle{\Delta}}/\tau})$.
Note that optical path imbalances between the BS and the detectors can be absorbed into simple offsets in the detector windows in time and do not affect the fidelity or yield.

The coherence factors $C_\gamma$ for each of the four heralded states in Eqs. \ref{eqn:state1}-\ref{eqn:state4} are
\begin{align}
C_{1001} & = \left(\frac{-2\mathfrak{t}^2\mathfrak{r}^2}{\mathfrak{t}^4+\mathfrak{r}^4}\right)\frac{1}{Y}\int_{\mu A'} dt_\mu \int_{\nu B'} dt_\nu I(t_\mu,t_\nu) \label{eqn:Copp}\\
C_{0110} &=  \left(\frac{-2\mathfrak{t}^2\mathfrak{r}^2}{\mathfrak{t}^4+\mathfrak{r}^4}\right)\frac{1}{Y}\int_{\mu B'} dt_\mu \int_{\nu A'} dt_\nu I(t_\mu,t_\nu) \label{eqn:Copp2}\\
C_{1100} &=  \hspace{17mm}\frac{1}{Y}\int_{\mu A'} dt_\mu \int_{\nu A'} dt_\nu I(t_\mu,t_\nu)\\
C_{0011} &=  \hspace{17mm}\frac{1}{Y}\int_{\mu B'} dt_\mu \int_{\nu B'} dt_\nu I(t_\mu,t_\nu),
\label{eqn:Csame}
\end{align}
where
\begin{align}
I(t_\mu,t_\nu)=f_{A}(t_\mu) f_{B}(t_\mu) f_{A}(t_\nu) f_{B}(t_\nu) \mathcal{M}(t_\mu,t_\nu)
\label{eqn:Integrand}
\end{align}
is integrated over the detection windows of the two relevant detectors for each type of correlation. 

If the BS has an power imbalance $\delta_{BS} \equiv \mathfrak{t}^2 - \mathfrak{r}^2$, we find that the fidelity $F^- \equiv F_{1001}=F_{0110}$ is degraded by a factor $(1-\delta_{BS}^2)/(1+\delta_{BS}^2)$, while the success probability $\mathcal{P}^-$ is boosted by a factor $1+\delta_{BS}^2$. 
For the case of detectors on the same side heralding the state $\ket{\Psi^+_{AB}}$, the BS imbalance does not affect the fidelity $F^+ = F_{1100}=F_{0011}$ but the success probability $\mathcal{P}^+$ is degraded by the factor $1-\delta_{BS}^2$.

The motional overlap factor in Eq. \ref{eqn:Integrand},
\begin{align}
\mathcal{M}(t_\mu,t_\nu) = \prod_{i} \mathcal{M}_{Ai}(t_\mu,t_\nu)\mathcal{M}_{Bi}(t_\mu,t_\nu)^*
\end{align}
has six terms representing the three directions of motion $i$ for each atom and can lead to decoherence of the atomic memories through the trace over the motion (Eq. \ref{eqn:trace}). For pure coherent states of motion $\ket{\beta_{qi}(t)}$ and $\ket{\beta^T_{qi}(t)}$, the motional overlap factors are
\begin{align}
\mathcal{M}_{qi}(t_\mu,t_\nu)&= \langle \beta^T_{qi}(t_{\nu}) \ket{\beta_{qi}(t_{\mu})}.  \label{eqn:overlap}
\end{align}

\CM{More generally, the initial state of motion of atom $q$ along direction $i$ appearing as $\alpha_{qi}$ in Eqs. \ref{eqn:alpha1}-\ref{eqn:alpha2} can be described by a density matrix distribution of coherent states, characterized by the Glauber P-distribution $P(\alpha_{qi})$ \cite{Glauber1963},
\begin{equation}
\rho^{M}_{qi}=\prod_{i}\int\mathrm{d}^2\alpha_{qi}|\alpha_{qi}\rangle\langle\alpha_{qi}| P(\alpha_{qi}). \label{eqn:Pdist}
\end{equation}
We take the initial state of motion as a thermal mixed state ${P(\alpha_{qi})=e^{-|\alpha_{qi}|^2/\bar{n}_{qi}}/(\pi\bar{n}_{qi})}$ \cite{Glauber1963}, which would be expected from laser-cooling. Here, $\bar{n}_{qi}$ is the mean thermal motional quantum number of atom $q$ in direction $i$. The thermal density matrix is then
\begin{equation}
\rho^{M,\text{Therm}}_{qi}=\prod_{i}\frac{1}{\pi\bar{n}_{qi}}\int\mathrm{d}^2\alpha_{qi}|\alpha_{qi}\rangle\langle\alpha_{qi}| e^{-|\alpha_{qi}|^2/\bar{n}_{qi}}, \label{eqn:thermal}
\end{equation}
Finally taking the trace over the motion and using Eqs. \ref{eqn:alpha1}-\ref{eqn:alpha2}}, we find
\begin{align} \label{eqn:M}
\mathcal{M}_{qi}(t_\mu,t_\nu)
& = \frac{1}{\pi\bar{n}_{qi}}\int d^2\alpha_{qi} \langle \beta^T_{qi}(t_{\nu}) \ket{\beta_{qi}(t_{\mu})}
e^{-|\alpha_{qi}|^2/\bar{n}_{qi}} \nonumber \\
&=e^{-i\psi_{qi}}e^{-(2\bar{n}_{qi}\!+\!1)Z_{qi}(t_{\mu},t_{\nu},T)}. 
\end{align}
%where in the last equation we assume ${(2\bar{n}_{qi}\!+\!1)Z_{qi}(t_{\mu},t_{\nu},T) \ll 1}$.% and $\psi_{qi} \ll 1$.

The decoherence term $Z_{qi}(t_{\mu},t_{\nu},T)$
characterizes the overlap of the recoiled motional states for atom $q$ and mode $i$ for $\bar{n}_{qi}=0$ (or any pure initial motional state), due to the random emission times $t_{\mu}$ and $t_{\nu}$ and the fixed time-bin separation $T$:
\begin{align} \label{eqn:D}
Z_{qi}&(t_{\mu},t_{\nu},T)= \tfrac{1}{2}|\beta^T(t_\nu)-\beta(t_\mu)|^2\\
%&=\tfrac{1}{2}|i(\eta'_{qi}-\eta_{qi}e^{i\omega_{qi}t'})e^{i\omega_{qi}T}-i(\eta'_{qi}-\eta_{qi}e^{i\omega_{qi}t'})|^2 \nonumber \\
&= \eta_{qi}'^2 (1-\cos\omega_{qi}T) 
+\eta_{qi}^2 [1-\cos\omega_{qi}(T+t_{\nu}-t_{\mu})]  \nonumber  \\
&\hspace{5mm}+\eta'_{qi}\eta_{qi} [\cos\omega_{qi}(T+t_{\nu})-\cos\omega_{qi}t_{\nu} \nonumber \\
&\hspace{15mm}+\cos\omega_{qi}(T-t_{\mu})-\cos\omega_{qi}t_{\mu}]. \nonumber \end{align}

The phase of the coherence appearing in Eq. \ref{eqn:M}  is independent of the initial motional state,
\begin{align} \label{eqn:phase}
 \psi_{qi} &= \eta_{qi}'^2\sin\omega_{qi}T + \eta_{qi}^2\sin\omega_{qi}(T+t_{\nu}-t_{\mu}) \\
 &-\eta'_{qi}\eta_{qi}\left[\sin\omega_{qi}(T+t_{\nu})+\sin\omega_{qi}t_{\nu} \right.\nonumber \\
 &\hspace{10mm}\left.+\sin\omega_{qi}(T-t_{\mu})- \sin\omega_{qi}t_{\mu}\right].\nonumber
\end{align}
This phase will fluctuate with the distributed photon detection times, but when averaged over the distribution of photon detection times, the resulting loss of contrast is much smaller than the main source of decoherence given by the real exponential term in Eq. \ref{eqn:M}, so we neglect it.

Assuming $\omega_{qi}\tau \ll 1$, the motional phase factors $\omega_{qi}t_{\pi}$ are small, simplifying the decoherence factor to
\begin{align} \label{eqn:Dapprox}
Z_{qi}(t_{\scriptscriptstyle{\Sigma}},t_{\scriptscriptstyle{\Delta}},T)=(\eta_{qi}-\eta_{qi}')^2(1-\cos&\omega_{qi}T)  \\
+ \tfrac{1}{4}\eta_{qi}'\eta_{qi}(1-\cos\omega_{qi}T)&\omega_{qi}^2 t_{\scriptscriptstyle{\Sigma}}^2 \nonumber \\
+\left[\tfrac{1}{4}\eta_{qi}\eta_{qi}'(1-\cos\omega_{qi}T)+\tfrac{1}{2}\eta_{qi}^2\cos\omega_{qi}T\right]&\omega_{qi}^2 t_{\scriptscriptstyle{\Delta}}^2. \nonumber
\end{align}
Here, $t_{\scriptscriptstyle{\Delta}}=t_{\nu}-t_{\mu}$ is the time difference between $\nu$ and $\mu$ photon detections with respect to their excitation pulses and $t_{\scriptscriptstyle{\Sigma}}=t_{\nu}+t_{\mu}$ is their sum. In the above expression, we have ignored terms that are odd in the random variable $t_{\scriptscriptstyle{\Delta}}$, as they will cancel when integrated over its symmetric distribution. 

The first and second terms in Eq. \ref{eqn:Dapprox} describe the recoil decoherence from the time-bin separation $T$ not in synchrony with the motional oscillation period and depends on both excitation and emission wavevectors. These terms vanish when $T$ is set to be commensurate with the oscillation periods of all 3 oscillators in both emitters \cite{Saha2025}. Taking ${\omega_{qi}T = 2\pi N_{qi} + \omega_{qi} \tilde{T}_{qi}}$, where $N_{qi}$ are integers and all six modes are nearly commensurate with the time-bin separation, or $\omega_{qi}\tilde{T}_{qi} \ll 1$, the decoherence factor takes the form
\begin{align}  \label{eqn:Dapprox2}
Z_{qi}(t_{\scriptscriptstyle{\Sigma}},t_{\scriptscriptstyle{\Delta}},T) &\approx \tfrac{1}{2}\left[(\eta_{qi}-\eta_{qi}')^2 + \tfrac{1}{4}\eta_{qi}'\eta_{qi}\omega_{qi}^2 t_{\scriptscriptstyle{\Sigma}}^2\right]\omega_{qi}^2\tilde{T}_{qi}^2 \nonumber \\
&+ \tfrac{1}{2}\eta_{qi}^2\omega_{qi}^2 t_{\scriptscriptstyle{\Delta}}^2.
\end{align}
The last term in Eqs. \ref{eqn:Dapprox} and \ref{eqn:Dapprox2} characterizes the recoil decoherence from random emission time differences $t_{\scriptscriptstyle{\Delta}}$\CM{, even when time-bin encoding is not employed}.

We can now integrate Eqs. \ref{eqn:Copp}-\ref{eqn:Csame} over the detection windows of $t_\mu$ and $t_\nu$ (or equivalently $t_{\scriptscriptstyle{\Sigma}}$ and $t_{\scriptscriptstyle{\Delta}}$), using Eqs. \ref{eqn:M} and \ref{eqn:Dapprox2}, and taking ${(2\bar{n}_{qi}\!+\!1)Z_{qi}(t_{\scriptscriptstyle{\Sigma}},t_{\scriptscriptstyle{\Delta}},T) \ll 1}$. 
%We finally present the central result: decoherence from recoil, reflected by the motional decoherence term from Eqs. \ref{eqn:Integrand}, \ref{eqn:M}, and \ref{eqn:Dapprox2}. 
%With no imbalances from the emitters and optical instrumentation, the time distributions of the two photons are the same: $f_A(t)=f_B(t)= e^{-t/2\tau}\Theta(t)/\sqrt{\tau}$. 
The Bell state fidelity is
\begin{align} \label{eqn:E1}
F_\gamma &= 1 - \sum_{qi}\frac{(\bar{n}_{qi}\!+\!\tfrac{1}{2})}{Y}
\iint_W dt_{\scriptscriptstyle{\Sigma}} dt_{\scriptscriptstyle{\Delta}}\frac{e^{-t_{\scriptscriptstyle{\Sigma}}/\tau}}{2\tau^2} Z_{qi}(t_{\scriptscriptstyle{\Sigma}},t_{\scriptscriptstyle{\Delta}},T) \\
&= 1 - \sum_{q}\left( \mathcal{E}^T_{q} + \mathcal{E}^R_{q} \right).  \label{eqn:E2}
\end{align}
The time-bin asynchronous error summed over all modes in Eq. \ref{eqn:E2} is
\begin{align}
\mathcal{E}^T_{q} &= \tfrac{1}{4}\sum_{i}(2\bar{n}_{qi}\!+\!1)(\eta_{qi}-\eta_{qi}')^2\omega_{qi}^2 \tilde{T}_{qi}^2\label{eqn:ET} \\
&= \tfrac{1}{4}\sum_i (2\bar{n}_{q i}\!+\!1)\omega^{\scriptscriptstyle{\Delta} R}_{qi}\omega_{qi} \tilde{T}_{qi}^2
\label{eqn:ET2}
\end{align} 
where the differential recoil frequency is given by ${\omega^{\scriptscriptstyle{\Delta} R}_{qi} = \hbar (k_{qi}-k'_{qi})^2b_{qi}^2/2m}$.
%=(\eta_{qi}-\eta'_{qi})^2\omega_{qi}}$. 
We have neglected the second term in the first line of Eq. \ref{eqn:Dapprox2}, which is only important when $\eta_{qi}\approx\eta_{qi}'$, or for co-linear excitation and emission directions, which is experimentally \CM{undesireable owing to large potential background counts or detector blinding from the excitation laser}.

Of course the error $\mathcal{E}^T_{q}$ can be eliminated by synchronizing the time-bins to be a multiple of all trap periods so that $\tilde{T}_{qi}=0$ \CM{as discussed above} \cite{Saha2025}. This may not always be practical for axial modes of a chain of trapped ions or \CM{atoms confined in} optical tweezers \cite{Li2024,Li2025}. In cases where the time-bin separation is desired to be much shorter than the period of atomic motion, the asynchronous time-bin separation \CM{should be set well larger than the atomic lifetime $\tau$ so that the two time-bins do not significantly overlap and result in indistinguishability of the photonic qubit states. For a time-bin separation set to $T=\tilde{T}_{qi}=\ell \tau$, we expect an additional fidelity error due to overlap of $e^{-\ell}$. In practice, the value of $\ell$ can be set to be much smaller than the motional decoherence error at the focus of this paper.}
%can be set as small as $T=\tilde{T}_{qi}=\ell \tau$ with $\ell$ large enough so that the time-bin overlap error $e^{-\ell}$ is negligible.

The random emission time error in Eq. \ref{eqn:E2}, \CM{relevant to any photonic qubit encoding (not just time-bin) is similarly}
\begin{align}
\mathcal{E}^R_{q} &= \sum_i (2\bar{n}_{qi}+1) 
\omega^R_{qi}\omega_{qi} \tau^2 W, \label{eqn:ER}
\end{align}
where the emission recoil frequency is given by ${\omega^R_{qi} = \hbar k_{qi}^2/2m}$. Here the parameter 
\begin{align}
W=\frac{1-(1+w+w^2/2)e^{-w}}{1-e^{-w}}
\end{align}
characterizes the variance of the random variable $t_{\scriptscriptstyle{\Delta}}$ over the detection window ${|t_{\scriptscriptstyle{\Delta}}| \le T_{\scriptscriptstyle{\Delta}}}$ (Fig. \ref{fig:integ}) and smoothly increases from $0$ to $1$ as the relative window size ${w = T_{\scriptscriptstyle{\Delta}}/\tau}$ increases from $0$ to $\infty$ \cite{Saha2025}.

\CM{We note that the time-bin encoding error depends on the recoil momentum transfer $\hbar(k_{qi}-k'_{qi})$ associated with the difference been excitation and emission, while the random emission error depends on the momentum transfer $\hbar k_{qi}$ associated only with emission. This is because in the absence of time-bin encoding, the excitation of the two atoms is simultaneous.}

\subsection*{Entanglement Recoil Errors for Common Atoms}

\CM{We now consider the asynchronous time-bin and random detection/emission errors under the assumption of Doppler laser cooling of the host atoms \cite{Leibfried2003}. The Doppler cooling limit results in thermal distribution of motional quanta with ${\bar{n}_{qi}=(2\omega_{qi}\tau)^{-1}}$, which is much greater than unity in typical experimental systems \cite{Leibfried2003, Saha2025, Li2024, Li2025}. By also taking ${k_{qi}\sim k}$ and $k_{qi}-k'_{qi}\sim k$, we find that the above errors simplify and become independent of the atomic motional frequency $\omega_{qi}$. 
We assume the time-bin asynchronous error is set to be faster than a single period of atomic motion as described in the previous section. We then find that the errors simplify to
\begin{align}
\mathcal{E}^{T,\text{Dopp}}_{q} &= \tfrac{1}{4}\ell^2\omega^{\scriptscriptstyle{\Delta}R}_{qi}\tau \label{eqn:ETsimple} \\
\mathcal{E}^{R,\text{Dopp}}_{q} &= W\omega^{R}_{qi}\tau.  \label{eqn:ERsimple}
\end{align}
For simplicity, we have collapsed the sums over motional modes in Eqs. \ref{eqn:ETsimple} and \ref{eqn:ERsimple} and replaced them with a single term when there are multiple motional-coupled atomic emitters,} as $b_{qi}^2 \sim 1/N$.

Both expressions above involve only atomic parameters and are summarized in Table \ref{table:errors} for common atomic systems under Doppler cooling. By employing sub-Doppler cooling to near the zero-point $({\bar{n}_{qi}=0})$ through resolved-sideband, EIT, or Sisyphus cooling for example, \CM{the recoil decoherence in Eqs. \ref{eqn:ETsimple}-\ref{eqn:ERsimple} and Table \ref{table:errors} are reduced further by the factor $\omega_{qi}\tau \ll 1$. This would result in entanglement errors being reduced by an order of magnitude for typical ion trap systems \cite{Saha2025} and by two or more orders for typical neutral atom systems \cite{Li2024,Li2025}.}

\begin{table}[t!]
\caption{Entanglement fidelity errors from recoil of two-photon heralded entanglement protocol due to random emission times, selecting events with $|t_{\scriptscriptstyle{\Delta}}|\le 2\tau$ ($w=2$, $W=0.38$ and $86\%$ yield). Recoil is taken to be along a single direction $i$ of motion for each \CM{of the two emitters} laser-cooled to \CM{their} Doppler limit. The time-bin \CM{separation $T=\tilde{T}_{qi}=\ell\tau$ is set as a multiple $\ell$ of the photonic width $\tau$ so that the photonic qubit indistinguishability error $e^{-\ell}$ is the same as the time-bin asynchronous recoil error of Eq. \ref{eqn:ETsimple}.}
%error assumes the time-bin separation is set so that $T=\tilde{T}_{qi}=\ell\tau$ with $\ell=4.6$ so that the overlap error is $1\%$. 
The errors are significantly smaller when the atoms are cooled to their zero point; the general formulas are given by Eqs. \ref{eqn:ET2}-\ref{eqn:ER}.}

\begin{tabular}{ |l|c|c|c|c|c| }
 \hline
    &transition & exc state & recoil  & time-bin  & random\\
    &wavelength  & lifetime & $\omega^R_{qi}/2\pi$    & error & error \\
\phantom{a}atom& $\lambda$ (nm) &  $\tau$ (ns)  & (kHz) & $2\mathcal{E}^{T,\CM{\text{Dopp}}}_{q}$ & $2\mathcal{E}^{R,\CM{\text{Dopp}}}_{q}$ \\
  \hline
    \hline
$^{\phantom{^{13}}9}\text{Be}^+$ & $313$ &  8.2 & $226$  & \CM{5.13}\%& $0.218\%$ \\ \hline
$^{\phantom{^{1}}40}\text{Ca}^+$ & $397$ &  6.8 & $31.6$ & \CM{1.28}\%& $0.025\%$ \\ \hline
$^{\phantom{^{1}}40}\text{Ca}^+$ & $866$ &  6.8 & $6.6$  & \CM{0.42}\%& $0.005\%$ \\ \hline
$^{\phantom{^{1}}87}\text{Rb}$   & $780$ &  26  & $3.8$  & \CM{0.74}\%& $0.012\%$ \\ \hline
$^{\phantom{^{1}}88}\text{Sr}^+$ & $422$ &  7.8 & $12.7$ & \CM{0.75}\%& $0.012\%$ \\ \hline
$^{\phantom{^{1}}88}\text{Sr}^+$ & $1092$&  7.8 & $1.9$  & \CM{0.18}\%& $0.002\%$ \\ \hline
$^{\phantom{^{1}}88}\text{Sr}$   & $461$ &  5.3 & $10.4$ & \CM{0.50}\%& $0.007\%$ \\ \hline
             $^{138}\text{Ba}^+$ & $493$ &  7.9 & $5.9$  & \CM{0.44}\%& $0.006\%$ \\ \hline
             $^{138}\text{Ba}^+$ & $650$ &  7.9 & $3.4$  & \CM{0.29}\%& $0.003\%$ \\ \hline
             $^{171}\text{Yb}^+$ & $369$ &  8.1 & $8.6$  & \CM{0.58}\%& $0.008\%$ \\ \hline
             $^{171}\text{Yb}$   & $399$ &  5.5 & $7.3$  & \CM{0.39}\%& $0.005\%$ \\ \hline
             $^{171}\text{Yb}$   & $1389$&330   & $0.6$  & \CM{1.22}\%& $0.023\%$ \\ \hline
\end{tabular}
\label{table:errors}
\end{table}

%********************************************************************
\subsection*{Active Cancellation of Recoil-Induced Decoherence}

We finally show how to use auxiliary operations to coherently eliminate the motional entanglement and thus \CM{potentially eliminate} the Bell state fidelity error stemming from the different recoil times of the emitters. \CM{This approach was recently considered in Ref. \cite{Apolin2025} and has some similarity to the use of special pulse-shaped optical drives that are used to coherently drive optical qubit transitions while avoiding decoherence from the residual entanglement with motion \cite{VanDamnitAll2025}.}

%Under synchronized time-bins $(\tilde{T}_{qi}=0)$ and 
After measuring both photons, the quantum state of the two atoms and their motion is ideally given by one of Eqs. \ref{eqn:state1}-\ref{eqn:state4} (taking $\mathfrak{t}^2=\mathfrak{r}^2 = 1/2$):
\begin{align} \label{eqn:entmotion}
\ket{\psi^{(3)}_\gamma} &= 
\frac{1}{\sqrt{2}}\ket{\downarrow_A\uparrow_B}\ket{\boldsymbol{\beta}_{A}(t_\mu)}\ket{\boldsymbol{\beta}_{B}^T(t_\nu)} \\ 
&\hspace{0.5mm}\pm\frac{1}{\sqrt{2}} \ket{\uparrow_A\downarrow_B} \ket{\boldsymbol{\beta}^T_A(t_\nu)}\ket{\boldsymbol{\beta}_{B}(t_\mu)}. \nonumber 
\end{align}
In order to disentangle the motion from the qubit states, we can apply particular state-dependent forces on the qubits, armed with the measured detection times $t_\mu$ and $t_\nu$ and the time-bin separation $T$. While the coherent states in Eq. \ref{eqn:entmotion} contain the initial (thermally distributed) motional states $\ket{\alpha_{qi}}$, the only remnant of the initial conditions after these rewinding operations is an overall phase factor, hence the final state is pure even after tracing over the motion. \CM{Such operations are very similar to the well-known M\o{}lmer-S\o{}rensen quantum gate operation \cite{MolmerPRL1999, SorensenPRL1999,Milburn2000ion,solano1999deterministic}, with several studies of errors in such operations \cite{SorensenPRA2000, Kirchhoff2024} and experimental demonstrations with fidelities greater than $99.9\%$ \cite{Harty2014,Ballance2016, Srinivas2021,Clark2021}. In practice, when such operations are implemented with optical sources, they require the atomic system to be confined to within the Lamb-Dicke limit, where $\eta_{qi}^2 \bar{n}_{qi} \ll 1$ \cite{SorensenPRA2000, Kirchhoff2024}, as is typically the case in ion trap experimental systems \cite{Leibfried2003}.}

As an example, by applying the state-dependent displacement operators 
\begin{align}
\hat{\mathcal{O}}_{qi}& = \ket{\downarrow_q}\bra{\downarrow_q}
\mathcal{D}(i\eta_{qi}e^{i\omega_{qi}t_\mu}-i\eta'_{qi}) \\
 &\hspace{0.5mm}+ \ket{\uparrow_q}\bra{\uparrow_q}
\mathcal{D}(i\eta_{qi}e^{i\omega_{qi}(t_\nu+T)}-i\eta'_{qi}e^{i\omega_{qi}T}) \nonumber
\end{align}
%\CM{comment on overall phase factor? Ken B. reference?}
to all relevant modes $i$ of both atoms $q$, the motion in Eq. \ref{eqn:entmotion} can be shown to disentangle from the qubits, independent of $\alpha_{qi}$. 
We find that any operator that provides a differential displacement to the two qubit states in each atom of 
\begin{align}
\Delta\alpha_{qi}& = i\eta_{qi}(e^{i\omega_{qi}t_\mu}- e^{i\omega_{qi}(t_\nu+T)})-i\eta'_{qi}(1-e^{i\omega_{qi}T}) \nonumber
\end{align}
for each motional mode $i$ will disentangle all modes of motion from the memories.
Schematically, these coherent rewinding operations reverse the excitation and emission recoil from each of the six modes (red and blue arrows in Fig. \ref{fig:phasespace}), returning the motion to its original state and thus eliminating motional decoherence entirely.
In practice, such forces can be realized through bichromatic optical Raman beams \cite{Monroe1996,Jia2022} or AC Stark shift gradients from optical tweezers \cite{Mazzanti2023}.

%********************************************************************
\CM{\subsection*{Outlook}
Quantum networks of atomic quantum memories are typically formed through the emission of photons. In this manuscript, we have undertaken a general study of potential fidelity-limiting effects involving the recoil of the atoms distributed over time. This includes the general random emission of photons from spontaneous emission, as well as special considerations for the case of time-bin photonic encoding. The resulting residual entanglement with atomic motion can lead to decoherence of the quantum memories at the network nodes. We have modeled this effect under realistic experimental conditions, and in the case of initially Doppler-cooled atoms, the result can be expressed in a very simple form that depends only upon fundamental atomic paramters. The levels of errors are less than $1\%$ and can be made even smaller with ground state cooling. Moreover, they can be reversed with appropriate post-entanglement quantum state-dependent forces common to popular ion trap quantum logic gates.
The results presented here agree with a recent experiment \cite{Saha2025} and are consistent with recent theoretical descriptions of similar phenomena involving emission into cavities \cite{Kikura2025} and light collection with large solid angles \cite{Apolin2025}.}

\vspace{5mm}

We acknowledge useful discussions with Jake Covey and Jeff Thompson. This work is supported by the DOE Quantum Systems Accelerator (DE-FOA-0002253) and the NSF STAQ Program (PHY-1818914). A.K. is supported by the AFOSR National Defense Science and Engineering Graduate (NDSEG) Fellowship.

\bibliographystyle{apsrev4-1}
%\bibliography{Refs}
%merlin.mbs apsrev4-1.bst 2010-07-25 4.21a (PWD, AO, DPC) hacked
%Control: key (0)
%Control: author (72) initials jnrlst
%Control: editor formatted (1) identically to author
%Control: production of article title (-1) disabled
%Control: page (0) single
%Control: year (1) truncated
%Control: production of eprint (0) enabled
%

\end{document}